\documentclass{IEEEtran4PSCC}
\ifCLASSINFOpdf
  \usepackage[pdftex]{graphicx}
\else
  \usepackage[dvips]{graphicx}
\fi
\usepackage[cmex10]{amsmath}
\usepackage{amsthm}
\usepackage{amsfonts}
\usepackage{amssymb}
\usepackage[dvipsnames]{xcolor}
\usepackage{array}
\usepackage{graphicx}
\usepackage{float}
\usepackage[colorinlistoftodos]{todonotes}
\usepackage{subfigure}
\usepackage{comment}
\usepackage{algorithm,algorithmic}
\makeatletter
\let\old@ps@headings\ps@headings
\let\old@ps@IEEEtitlepagestyle\ps@IEEEtitlepagestyle
\def\psccfooter#1{%
    \def\ps@headings{%
        \old@ps@headings%
        \def\@oddfoot{\strut\hfill#1\hfill\strut}%
        \def\@evenfoot{\strut\hfill#1\hfill\strut}%
    }%
    \def\ps@IEEEtitlepagestyle{%
        \old@ps@IEEEtitlepagestyle%
        \def\@oddfoot{\strut\hfill#1\hfill\strut}%
        \def\@evenfoot{\strut\hfill#1\hfill\strut}%
    }%
    \ps@headings%
}
\makeatother

\psccfooter{%
        \parbox{\textwidth}{\hrulefill \\ \small{21st Power Systems Computation Conference} \hfill \begin{minipage}{0.2\textwidth}\centering \vspace*{4pt} \includegraphics[scale=0.06]{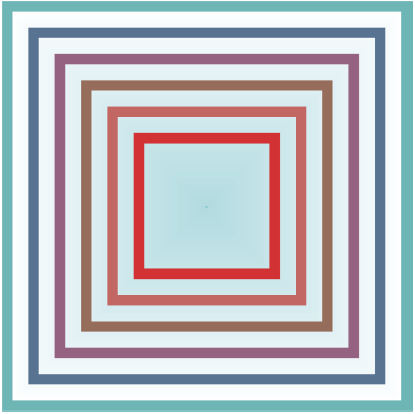}\\\small{PSCC 2020} \end{minipage} \hfill \small{Porto, Portugal --- June 29 -- July 3, 2020}}%
}

\begin{document}
\title{Probabilistic Analysis of Masked Loads with Aggregated
  Photovoltaic Production}

%% To specify the authors when (number of affiliations <= 2)
\author{
\IEEEauthorblockN{Shaohui Liu\\}
\IEEEauthorblockA{Department of Electrical and Computer Engineering  \\
The University of Texas at Austin\\
Austin, TX, USA\\
shaohui.liu@utexas.edu}
\and
\IEEEauthorblockN{Daniel Adrian Maldonado\\ Emil M. Constantinescu}
\IEEEauthorblockA{Mathematics and Computer Science \\
Argonne National Laboratory\\
Lemont, IL, USA\\
\{maldonadod, emconsta\}@anl.gov}
}

%% To specify the authors when (number of affiliations > 2)
% \author{\IEEEauthorblockN{Author n.1\IEEEauthorrefmark{1},
% Author n.2\IEEEauthorrefmark{2},
% Author n.3\IEEEauthorrefmark{3}, 
% Author n.4\IEEEauthorrefmark{3} and
% Author n.5\IEEEauthorrefmark{4}}
% \IEEEauthorblockA{\IEEEauthorrefmark{1} Department Name of Organization A\\
% Name of the organization A,
% Address A\\ Emails if wanted}
% \IEEEauthorblockA{\IEEEauthorrefmark{2} Department Name of Organization B\\
% Name of the organization B,
% Address B\\ Emails if wanted}
% \IEEEauthorblockA{\IEEEauthorrefmark{3} Department Name of Organization C\\
% Name of the organization C,
% Address C\\ Emails if wanted}
% \IEEEauthorblockA{\IEEEauthorrefmark{4}Department Name of Organization D\\
% Name of the organization D,
% Address D\\ Emails if wanted}
% }

\maketitle

\begin{abstract}In this paper we present a probabilistic analysis framework
  to estimate behind-the-meter photovoltaic generation in real time. We develop a forward model
  consisting of a spatiotemporal stochastic process that represents the photovoltaic generation and a stochastic differential
  equation with jumps that represents the demand. We employ this model to disaggregate the behind-the-meter photovoltaic generation using net load and irradiance measurements.
\end{abstract}

% Use this to place sponsorships
\thanksto{This material is based upon work supported by the
U.S. Department of Energy, Office of Science, Advanced Scientific
Computing Research under Contract DE-AC02-06CH11357.}

\section{Introduction}
\label{sec:introduction}

The increase in penetration of user-sized distributed energy resource (DER) systems poses
challenges for the planning and operation of the grid. A major issue
with behind-the-meter (BTM) solar generation is the lack of direct measurements of the instantaneous power injections. Furthermore, the volatility of solar generation production caused by 
weather variability (e.g., cloud coverage) brings additional uncertainty to forecasts \cite{Widen_2017a}.
This lack of observability makes  it difficult to quantify the aggregated
effect of BTM photovoltaics (PV) generation on the transmission grid.  Proper characterization of BTM PV generation in real time would allow utilities to prepare for  and quantify the risk of situations in which sudden ramps in generation occur or in which a large amount of DER systems trip after a fault. 

\iffalse
For the purpose of clarity and consistency, we first introduce the following terms:

\begin{itemize}
\item{\textit{PV generation:}} sum of all BTM PV power injections.
\item{\textit{Masked load:}} sum of all individual customer demand. With the increase of BTM-DER, PV generation "masks" the actual load consumption.
\item{\textit{Net load:}} masked load minus the distributed generation. This quantity, plus feeder losses, is measured at the substation.
\end{itemize}
\fi

Given the constraints that utilities face with respect to measurements,
forecasts, and parameters of inverters; recent studies have investigated
the issue of ``disaggregating" the PV power signal from the measurements
or inferring the instantaneous PV power through irradiance measurements
and other proxy regressors. One can consider the disaggregation
problem as an approximate algebraic relationship:
\begin{align}
\label{eq:power_balance}
P_{\textrm{NET}} = P_{\textrm{MASKED}} - P_{\textrm{PV}} \,
\end{align}
where $P_{\textrm{NET}}$ is the net power seen by the transmission grid (also called measured power),  $P_{\textrm{PV}}$ is the aggregated power generated by the distributed PV inverters, and $P_{\textrm{MASKED}}$ is the actual aggregated load demand that remains \textit{masked} by the PV production. In a high loading and high PV production scenario, the utility might underestimate the actual load in the feeder, and a voltage transient that trips a large amount of PV inverters may jeopardize the dynamic stability of the system.

Several researchers are investigating how to disaggregate the masked
load from the net load signals. Vrettos et al. \cite{vrettos_2019}
characterize the literature in three main groups: transposition model
approaches, data-driven approaches, and hybrid approaches. The
transposition methods involve extrapolating irradiance to a set of
inverter models to compute the PV generation. For instance, Engerer
and Mills~\cite{Engerer2014} use proxy measurements from a PV inverter
together with the clear-sky index and a PV inverter performance model
to extrapolate the generation of the rest of the inverters. In
\cite{killinger2016projection}, Killinger et al. further delve into
the cases in which the parameters of the PV inverters are not uniform,
and they develop a ``projection method" to calculate the global
horizontal irradiance (GHI) using a proxy power measurement. The GHI
measures the total amount of irradiance received by a flat surface
at the ground from above, and is a central indicator of the solar energy
that can be produced by PV panels. With regard to
the data-driven methods, Sossan et al. \cite{sossan_2018} and Patel et
al.~\cite{Patel2018} analyze the impact of global horizontal
irradiance fluctuations on the time series and use this
information to desegregate the PV generation from the net load
signal. The hybrid approaches include \cite{landelius_2018} and
\cite{Bright2018}. In the former publication, neural networks together
with load forecasts and PV production models are used to forecast the
net load. In the latter, Bright et al. use satellite-derived GHI
estimates in 10-minute intervals, together with PV generation models,
to interpolate to the aggregated PV generation. 

In this paper we propose a novel algorithm for the disaggregation of
instantaneous PV generation in a feeder that falls into the category of hybrid methods. Our methodology differs from
previous work in that we consider high-frequency measurements of irradiance
and net load. It has been shown in \cite{roberts_2016} and \cite{gudrun} that both the load and the PV generation, in short time intervals, can be characterized as stochastic processes with certain properties.

To extract information from the higher-resolution time series, 
 we fit statistics such as temporal variance, autocorrelation 
 and variogram,  which
allow us to obtain the parameters of the underlying masked load
process, provided that we have a model for the spatial irradiance and
the installed PV panels. While  modeling
the instantaneous PV generation with irradiance data and inverter
parameters can be error-prone, by leveraging the spatiotemporal
statistics produced by the irradiance fluctuation we can tolerate
higher errors in the model. To  this end, we develop a model for the geographically distributed PV
aggregation power injection with limited solar irradiance
measurements. We then design a modified stochastic differential
equation model based on the Ornstein-Uhlenbeck process with jumps to simulate the
masked load yielding a jump-diffusion process \cite{roberts_2016}. The
net load model then results from the combination of the two submodels. We further design a disaggregation algorithm to mitigate the error of masked load estimation caused by the estimated PV generation.
Simulation studies with
real recorded solar irradiance data and load data recorded by $\mu$PMU show that the spatiotemporal
model with the disaggregation algorithm is a tenable method to reduce disaggregation error. Moreover,
simulations indicate that we can accurately estimate the aggregated PV
active power generation at a distribution feeder with limited sensor
deployment. 

The rest of the paper is organized as follows. In section
\ref{sec:pv_power_model} we model the PV power with a novel spatial
Gaussian process (GP) for predicting solar irradiance under limited
observations. In section \ref{sec:ornstein_uhlenbeck_process} we model
the load power by an Ornstein-Uhlenback (OU) process with jumps. In section
\ref{sec:power_disaggregation} we propose a disaggregation algorithm
that separates the net power with real-time solar irradiance. In
section \ref{sec:conclusion_and_future_work} we concluded our work and
discussed the possible future work.

%%%%%%%%%%%%%%%%%%%%%%%%%%%%%%%%%%%%%%%%%%%%%%%%%%%%
% Spatial GP
%%%%%%%%%%%%%%%%%%%%%%%%%%%%%%%%%%%%%%%%%%%%%%%%%%%%

\section{Solar Generation Model}
\label{sec:pv_power_model}

In this section we consider the construction of a stochastic model of the aggregated PV power generation in a region, using sparse irradiance 
measurements and inverter performance equations. The irradiance
GP-based forecast follows a standard kriging framework with examples
that more recently include \cite{aryaputera2015}. 

\subsection{Spatial Gaussian Process for Clear Sky Index}

Whereas direct measurement of the instantaneous power injection of
each PV inverter is infeasible, one can build approximate 
models that simulate the spatial distribution of the irradiance and, 
together with data and models of the installed PV inverters, approximate
the total PV generation in a feeder. Thus, we focus our efforts on
developing a reliable model for the prediction of the aggregated
irradiance that acknowledges the sparsity of measurements. 

To build this model we employ a spatial Gaussian process to 
represent the variability  and spatial correlations in solar
irradiance.  Although the normal marginals do not represent well the
solar distribution, which tends to be bimodal (i.e., have two
concentration peaks corresponding to cloudy and sunny conditions \cite{Widen_2017b}), the
correlations are useful in determining how the irradiance in the
geographic region co-varies. This relation is expressed through the
conditional distribution, which is the critical ingredient in our
predictive framework. GPs have closed forms for the posterior 
and conditional distributions and this confers a distinctive advantage
in achieving fast simulation and sampling, which can be critical in real-time applications. We expect
 this conditional distribution to depend on the weather
conditions, season, and climate. Such a GP calibration process likely
needs to take place with varying degrees in each region where it is
deployed. In the next
step, the solar 
irradiance is used to estimate the PV power by propagating the
irradiance through a set of inverter models whose location and
parameters are assumed  
to be known. For simplicity, we will assume the parameters of these
inverters are uniform, and we neglect model errors. One approach to
alleviate this restriction is discussed in \cite{killinger2016projection}. 

We consider a realistic setting that assumes we have one or two global horizontal irradiance (GHI) observations per neighborhood. These observations are used to estimate
the total amount solar generation for the entire area.

The procedure we
used to estimate the forecast solar production is as follows. First we
measured the global horizontal irradiance, $G$, for several spatial
locations and calculated the clear-sky horizontal irradiance, $G_c$, for
the same locations. With these quantities we estimated the clear-sky
index, $\kappa$, for each site by 
\begin{align}
G = \kappa G_c \,.
\label{eq:csi}
\end{align} 
The clear-sky index represents the fraction of irradiance that passes
through atmosphere relative to clear-sky conditions. The advantage of
using the clear-sky index is that it is a detrended quantity. We assume a joint
distribution for $\kappa \sim \mathcal{N}(\mu,\Sigma)$, where the mean
$\mu$ is set to zero by debiasing the data, and the covariance matrix $\Sigma$ is a symmetric
positive definite matrix, $\Sigma = 
\left[\epsilon_{ij}\right]_{1\leq i,j\leq n}$. Many  models
for the covariance function exist. Here we employ a relatively simple
anisotropic kernel: 
\begin{align}
\epsilon_{ij} = \alpha \cdot \exp{\left(-\left(\theta^2_x (r^x_{ij})^2 +
 \theta^2_y (r^y_{ij})^2 \right)\right)} + \beta \cdot \delta_{ij}\,,
\label{eq:cov1}
\end{align}
with $i,j=1,\, 2,\,\dots,n$, where $\alpha$, $\beta$, $\theta_x$, and
$\theta_y$ are parameters; $r^x_{ij}$ and $r^y_{ij}$ are spatial
distances between site $i$ and site $j$ in the $x$ and $y$ directions,
respectively; $\delta_{ij}$ is the the Kronecker delta function; and
$\beta \cdot \delta_{ij}$ has the effect of a statistical nugget.  We
estimate the kernel parameters by a least-squares method:
\begin{align}
\min_{\alpha, \beta, \theta_x, \theta_y} || \Sigma_{model} - \Sigma_{obs} ||_2 \,,
\label{eq:pv_ls}
\end{align}
where $\Sigma_{obs}$ is the empirical covariance of measured $\kappa$ and
$\Sigma_{model}$ is given by the parametric function
\eqref{eq:cov1}. We note here that this is a spatial model aimed at
characterizing the irradiance variability in a small area. Because our
final measure in this study is total solar irradiance, we argue that
this model has sufficient complexity as our numerical experiments
illustrate. 

We have also implemented a maximum likelihood
estimation procedure; however, for our setup the differences were
negligible. The anisotropy, measured as the difference
between latitudinal (north-south) and longitudinal (east-west)
components of the GP kernel, seems to play an important role. We
recorded a difference of about 20\% between these components in our
experiments, which represents a point of departure from studies such
as \cite{aryaputera2015}. This indicates a predominant flow direction,
which confers more accurate predictions in space. 

We use the joint distribution to infer the clear sky index  at
unobserved locations. If we denote by $X_1$ the unobserved locations
and the observation sites $X_2$ then their joint distribution is
represented by
\begin{align}
  \label{eq:joint:GP}
\begin{bmatrix}
X_1 \\
X_2
\end{bmatrix}
\sim
\mathcal{N} \left(
    \begin{bmatrix}
\mu_1 \\
\mu_2
\end{bmatrix}
,\begin{bmatrix}
\Sigma_{11} & \Sigma_{12} \\
\Sigma_{21} & \Sigma_{22}
\end{bmatrix}\right) \,,
\end{align}
where $\Sigma_{ij}$ are block covariance matrices as in
\eqref{eq:cov1}. 
We compute the conditional distribution of $X_1$ provided observations
$X_2$: $(X_1|X_2)\sim\mathcal{N}(\mu',\Sigma^{'})$ and expressed in
closed form by
\begin{subequations}
  \label{eq:conditional:GP}
\begin{align}
\mu' & = \mu_1 - \Sigma_{12}\Sigma_{22}^{-1}(X_2 - \mu_2) \,, \\
\Sigma' &= \Sigma_{11} - \Sigma_{12}\Sigma_{22}^{-1}\Sigma_{21} \,.
\end{align}
\end{subequations}

\begin{figure}[!h]
\centering
%\hspace{-15mm}
\includegraphics[width=0.9\linewidth]{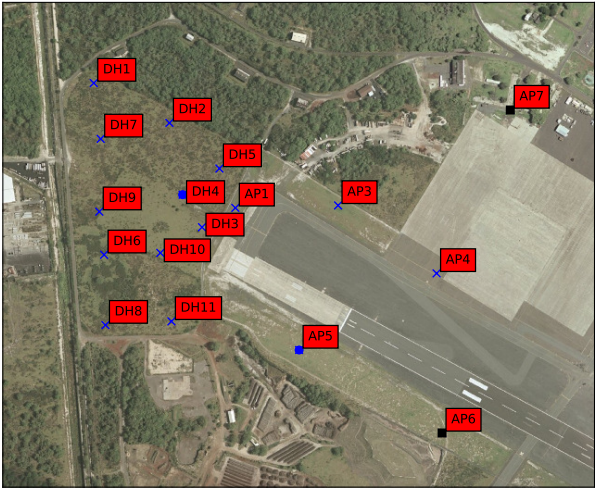}
\caption{Illustration of the solar PV and observation
  sites. The blue circles (DH4 and AP5) indicate stations that are used as
  observations. Black squares indicate remote sites that show smaller
correlation with the rest of them.}
\label{fig:solar:sites}
\end{figure}

We use a dataset that provides one-year's worth of global horizontal
solar irradiance data of 17 distributed sites collected every second
during daylight located on Oahu island, Hawaii
\cite{Sengupta_2010}. The clear-sky irradiance, $G_c$, is
calculated by pvlib \cite{holmgren2018pvlib} using the precise time
and coordinates data of solar panels assumed to be collocated with the
irradiance measurements. The site location and names are illustrated in Fig.~\ref{fig:solar:sites}. Indicated also are  two sites used as
observations and two that represent outliers in terms of proximity to
the site clusters. 

\begin{figure}[!h]
\centering
\includegraphics[width=1.0\linewidth]{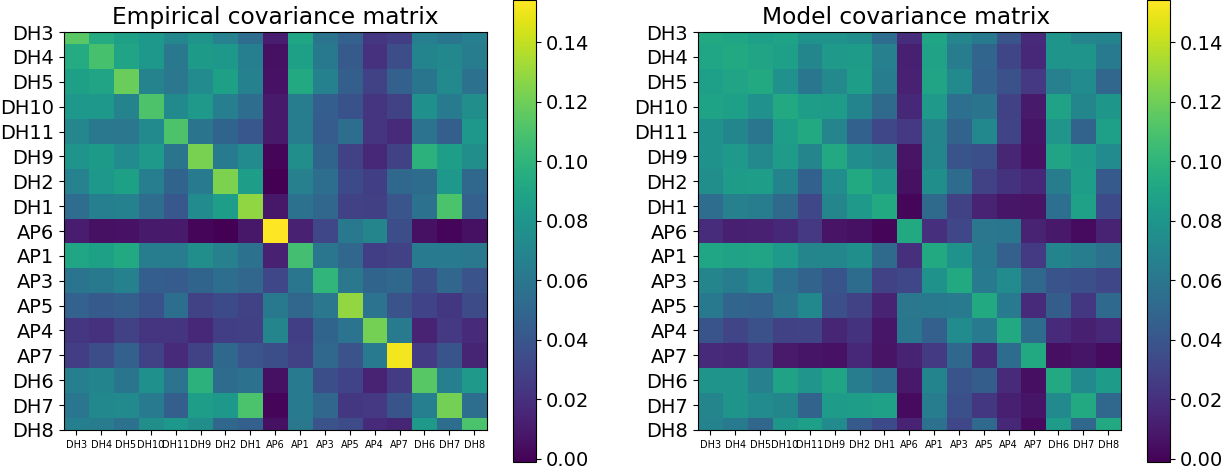}
\caption{The covariance matrix calculated by real $\kappa$ data and
  the  covariance matrix model calculated by using kernel \eqref{eq:cov1}.}
\label{fig:solar:cov_pred}
\end{figure}

We compute the empirical covariance of the 17 sites and plot the
matrix entries and the entries of the calibrated covariance
model \eqref{eq:cov1} in Fig.~\ref{fig:solar:cov_pred}. The covariance
represents how the 17 stations covary at every time instance in
space. We note that the model covariance approximates well the
structure of the empirical covariance. Moreover, we can see the
outlier locations being less correlated with the rest of them.   

\subsection{Power Model of PV Systems}
The following are derivations of the power model of PV system based on solar irradiance.

\begin{table}[h!]
\centering
\begin{tabular}{l l} 
 \hline
 Notation & Meaning  \\ [0.5ex] 
 \hline\hline
 $\kappa$ & clear-sky index  \\ 
 $G_c$ & clear-sky irradiance \\
 $k_d$ & diffuse fraction: $\frac{\textnormal{global\, irradiance}}{\textnormal{extraterrestrial \,irradiance}}$ \\
 $R_b$ & geometric factor: scaling factor of incidence angle \\
 $A_i$ & anisotropy index: $\frac{\textnormal{beam \, radiation}}{\textnormal{extraterrestrial\, radiation}}$ \\
$\beta$ & tilt angle of the tilted plane \\
$\rho_g$ & albedo of the ground \\
$A$ & total area of the PV array \\
$\eta$ & PV module conversion efficiency \\
$q_a$ & additional module/array loss \\
$P_{ac0}$ & rated max AC power of inverter \\
$P_{dc0}$ & DC power at which inverter reaches AC rating \\
$P_{s0}$ & inverter threshold power (start to give AC power)\\ [1ex] 
 \hline \\
\end{tabular}
\caption{Notation in PV Model and Definition}
\label{table:1}
\end{table}

GHI (global horizontal irradiance):
\begin{align}
G = \kappa G_c \,.
\end{align}

Diffuse irradiance and beam irradiance:
\begin{align}
G_d &= k_d G \,,\\
G_b &= G - G_d \,.
\end{align}

Global irradiance on tilted plane:
\begin{align}
\begin{split}
G_T &= G_bR_b + G_d\left((1-A_i)\frac{1+\cos\beta}{2} + A_iR_b\right)  \\
&+ G\rho_g\frac{1-\cos\beta}{2} \,.
\end{split}
\end{align}

%DC power output of a PV array in the tilted plane:
%\begin{align}
%P_{dc} = AG_T \eta(a-q_a) \,.
%\end{align}

AC power output is calculated as in \cite{Widen_2017a,holmgren2018pvlib} %(into the grid, Sandia inverter model):
\begin{align}
P_{ac} = P_{ac0}\frac{P_{dc} - P_{s0}}{P_{dc0} - P_{s0}} \,.
\label{eq:ac_power}
\end{align}

We designed an experiment on synthetic data involving 17 sites with
2 observed sites. The irradiance data was sampled at $1$Hz frequency,
which is the same as with commonly used sensors. 

\begin{figure}[!ht]
\centering
\hspace{-0mm}\includegraphics[width=\linewidth]{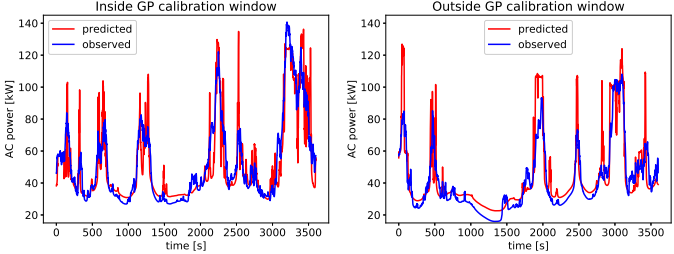}\\
%\hspace{-0mm}\includegraphics[width=1.1\linewidth]{Figures/PV_power_fcst.eps}
 %where an .eps filename suffix will be assumed under Latex, 
 %and a .pdf suffix will be assumed for pdfLatex; or what has been declared
 %via \DeclareGraphicsExtensions.
\caption{Aggregate AC solar power observed and predicted by
  using two observation sites using eq. (\ref{eq:ac_power}). The forecast correspond to April 8,
  2010, (left) 11 am to noon local time where three sites were used for
  forecast, and (right) noon to 1 pm where the same GP fit was used to
  make predictions.} 
\label{fig:solar:power}
\end{figure}

Numerical results suggest
that the GP model can precisely recover the covariance matrix with
limited observations. Furthermore, the forecast method  predicts
the PV power production of the aggregated sites without complete
observations, except for when sharp jumps in the irradiance
are caused by clouds moving in or our of the area.
In Fig.~\ref{fig:solar:power} we illustrate the aggregated PV power
computed by collocating uniform PV panels with deployed GHI sensors
(Fig.~\ref{fig:solar:sites}), which is referred to as the observed PV
power. We also use the GP procedure to estimate the clear sky index
based on two observation sites for the entire region,
compute the irradiance, and use the same PV model to estimate the PV power. 
The joint GP process \eqref{eq:joint:GP} is calibrated by using data
that corresponds to the time frame in
Fig.~\ref{fig:solar:power} (left). The same GP process is used to make
prediction corresponding to the time frame in
Fig.~\ref{fig:solar:power} (right). In other words, the GP is
calibrated with the entire sensor network data for
Fig.~\ref{fig:solar:power} (left), whereas in the second case the GP
has  access only to the designated observed sites. These results
indicate a relatively good prediction capability of the spatial
GP. The least accurate predictions are likely associated with sudden
irradiance jumps linked to the incoming of clouds through an unobserved
section of the PV generators. Extensions to temporal models might help
alleviate these aspects, as illustrated for short forecast-time ahead
in \cite{aryaputera2015,Bessac_A2018}.  

%%%%%%%%%%%%%%%%%%%%%%%%%%%%%%%%%%%%%%%%%%%%%%%%%%%%
% OU
%%%%%%%%%%%%%%%%%%%%%%%%%%%%%%%%%%%%%%%%%%%%%%%%%%%%

\section{Ornstein-Uhlenbeck Process for Load Modeling}
\label{sec:ornstein_uhlenbeck_process}

We model the masked load as an Ornstein-Uhlenbeck process with spikes following \cite{roberts_2016}: 
\begin{align}
dx_t = \gamma(\mu-x_t)dt + \sigma d\omega_t + J_tdq_t ,
\label{eq:ou_sde}
\end{align}
where $\gamma$ is the mean reversion rate and $\mu$ the long-term mean
of the OU process. The variance of the OU process represents normal
load changes such as small loads being switched on and off, and larger
spikes represent sudden and less frequent switching of larger
loads. We model $d\omega_t$ as a a standard Wiener process with
diffusion $\sigma$ \cite{Gardiner_B1985}. For the jumps, we follow a Poisson process, where
$J_t$ is a random variable,  $|J_t|$ follows Gamma distribution, and
$q_t$ is the Poisson random variable with intensity $\lambda$: 
\begin{align}
dq_t = \begin{cases}
&1, \,\textnormal{w/ \, probability} \, \lambda dt \\
&0, \, \textnormal{w/ \, probability} \, 1-\lambda dt \,.
\end{cases}
\end{align}

\subsection{Numerical Solution}

We discretize (\ref{eq:ou_sde}) and derive the first-order numerical solution by
the Euler-Maruyama method with step $\Delta t$:
\begin{align}
x_{i+1} = x_i + \gamma(\mu-x_i)\Delta t + \sigma(W_i - W_{i-1}) + J_i (P_i - P_{i-1})
\label{eq:ou_em}
\end{align}
where $\Delta W_i =W_i - W_{i-1}$ and $\Delta W_i\sim \sqrt{\Delta t} \,\mathcal{N}(0,1)$
are the independent increments. In particular, the model \ref{eq:ou_sde} has an explicit solution form that could be discretized as
\begin{align}
x_{i} = \mu &+ (x_0 - \mu)e^{-i\Delta t\gamma} + \sigma \sum_{j=1}^i e^{-\gamma(i-j+1)\Delta t}\cdot (W_j - W_{j-1}) \nonumber\\
&+ \sum_{j=1}^i e^{-\gamma(i-j+1)\Delta t}\cdot J_{i-1}(q_i - q_{i-1}).
\label{eq:ou_direct}
\end{align}
Note that our stochastic differential equation (SDE) model is for a relatively smooth system. Thus the numerical error for the Euler Maruyama scheme is relatively small, and the two methods generate nearly identical numerical solutions in our test cases. Furthermore, the computational cost of numerical scheme \eqref{eq:ou_em} is $O(n^2)$, while the cost of exact solution \eqref{eq:ou_direct} is $O(n^3)$ with respect to the number of steps. So we implement the Euler Maruyama  scheme in our algorithm.
\subsection{Parameter Estimation}

For parameter estimation with the discrete time series $\{X_i\}_{i=0}^{N}$ we first consider a simple filter. We let $y_{i+1} := f(X_i)$, where $f(X_i) = X_i + \gamma(\mu - X_i)\Delta t$. Then:
\begin{align}
X_{i+1} = y_{i+1} + \xi_{i+1},
\label{eq:ou_filter}
\end{align} 
where $\xi_i \sim \mathcal{N}(0,\sigma^2\Delta t)$ if we do not consider the rare jumps at first. With this preprocessing, we have the following algorithm.
\begin{algorithm}
\caption{OU Parameter Estimation}
\label{alg:ou_parameter_estimation}
\begin{algorithmic}
\STATE \textbf{Input:} PMU data $\{x_i\}_{i=0}^N$
\STATE $\cdot$Estimate the mean reversion rate $\gamma$ by martingale function
\STATE $\cdot$Calculate the random process set $\{\xi_i\}_{i=0}^N$ by \eqref{eq:ou_filter}
\STATE $\cdot$Calculate the mean $\mu_0$ and variance $\sigma_0$ of $\{\xi_i\}$
\FOR {$1\leq i \leq N$}
        \STATE $\cdot$Identify jumps $\{J_j\}$ by $3\sigma_0$
\ENDFOR
\IF {$\# jumps > 0$} 
        \STATE $\cdot$Calculate Poisson parameter $\lambda$
        \STATE $\cdot$Estimate Gamma parameters shape $k$, scale $\theta$ for $\{J_j\}$
\ELSE
        \STATE $\cdot$No jumps identified
\ENDIF 
\STATE $\cdot$ Calculate the modified increment mean $\mu_1$ and variance $\sigma_1$ for $\{\xi_i\}_{i=0}^N \backslash \{J_j\}$
\STATE $\cdot$  Run Kolmogorov-Smirnov test for Gaussian and Gamma 
\IF {KS test passed}
        \STATE \textbf{Return:} $\mu_1,\sigma_1,\lambda,k,\theta$ 
\ENDIF
\end{algorithmic}
\end{algorithm}

We used an unbiased method based on a martingale estimation function to estimate the mean reversion rate $\gamma$ \cite{roberts_2016}. In particular, the estimator is unbiased, consistent, and asymptotically normally distributed given the assumption that the underlying diffusion in the SDE model is ergodic~\cite{martingale_1995}. We first write the martingale estimation function as
\begin{align}
G_N (\gamma) = \sum_{i=1}^N \frac{\dot{b}(x_{i-1};\gamma)}{\sigma_{i-1}^2}\{x_i - \mu_t - (x_{t-1} - \mu_{t-1})e^{-\gamma}\},
\label{eq:martingale}
\end{align}
where
\begin{align}
b(x_t;\gamma) = \frac{d\mu_t}{dt} + \gamma(\mu_t - x_t).
\end{align}
The estimation of the $\gamma$ is the unique zero point of  \eqref{eq:martingale}:
\begin{align}
\hat{\gamma} = -\log\left(\frac{\sum_{i=1}^N Y_{i-1}\{x_i - \mu_i\}}{\sum_{i=1}^N Y_{i-1}\{x_{i-1} - \mu_{i-1}\}}\right),
\label{eq:gamma_estimator}
\end{align}
where
\begin{align}
Y_{i-1} = \frac{\mu_{i-1} - x_{i-1}}{\sigma_{i-1}^2}.
\end{align}

With the parameter estimation Algorithm \ref{alg:ou_parameter_estimation}, we can estimate the parameters for the OU process and then generate the predictions using the numerical solution scheme \eqref{eq:ou_em}. 

%%%%%%%%%%%%%%%%%%%%%%%%%%%%%%%%%%%%%%%%%%%%%%%%%%%%
% BTM PV
%%%%%%%%%%%%%%%%%%%%%%%%%%%%%%%%%%%%%%%%%%%%%%%%%%%%

\section{BTM PV Generation Disaggregation}
\label{sec:power_disaggregation}

In real cases, after the installment of PV panels in the grid, we no longer have direct measurements of the load power. The only accessible data is net power measured by $\mu$PMU and limited measurements of solar irradiance. Thus we need to reduce the uncertainties for more precise prediction and planning. As with the separate tests in PV power and load power, the data set of GHI is in $1$ Hz for 10 minutes, and we also down sampled the $\mu$PMU data to $1$ Hz for 10 minutes for convenience.

\iffalse
\subsection{Solar Irradiance and PV Power}

 Specifications of the PV panels follow the parameters in \cite{Widen_2017a}. 
\begin{figure}[!ht]
\centering
\hspace{+0mm}\includegraphics[width=0.8\linewidth]{PV_power_pred.eps}
\caption{The AC power calculated by full irradiance data v.s. the AC power forecast. The forecast corresponds to April 8,
  2010, 11am-noon local time where the GP fit was used to
  make predictions based on observations on two sites.}
\label{fig:solar:pv_power_pred}
\end{figure}

Experiments (Fig. \ref{fig:solar:pv_power_pred}) suggest that given 2 sensors installed at a 17 distributed PV panel arrays, our GP model can predict the aggregated PV power production accurately and timely, especially for the sharp jumps caused by the sudden changes of the local weather due to the movements of clouds. However, our current GP model is still sensitive to the jumps of weather, and thus we are unable to precisely predict the scale of sudden jumps. This is caused by the rank-deficiency of the input data, and we are still making improvements to it {\textcolor{red}{need further discussion}}.
\fi

\subsection{Net Load Model}
We consider the linear power model of the distributed system with
three components: the net load - external power injection, the masked
load - the sum of the consumer demand and the aggregated BTM PV power
within the system \eqref{eq:power_balance}. 
%\begin{align}
%P_{NET} = P_{\rm ML} - P_{\rm PV}
%\label{eq:net_load}
%\end{align}
In real cases, while we have direct measurements on the net load, we have no direct information about the PV power and masked load power. PV power production could be predicted by the partial information of the solar irradiance. Moreover, we can estimate the corresponding masked load based on the estimate of BTM PV power and make further predictions. Thus the key to accurate prediction is a reliable disaggregation algorithm using limited information.

\subsection{Disaggregation Strategy\label{sec:Disaggregation:Algrithm}}

For the Ornstein-Uhlenbeck process with jumps, we consider a full parameter vector:
\begin{align}
\Theta = \left[ \gamma, \mu, \mu_1, \sigma_1, k, \theta, \lambda \right]
\label{eq:ou_param_vec}
\end{align}
where $\gamma, \mu$ are the mean reversion rate and the long-term mean
of OU process; $\mu_1,\sigma_1$ are the mean and standard deviation of
Wiener process; $k,\theta$ are the shape and scale parameters of the Gamma
distribution that describes jumps; and $\lambda$ is the parameter for the
Poisson process.

Then we can take the OU parameters calculated by
Algorithm \ref{alg:ou_parameter_estimation} using the $\mu$PMU data recorded
before the installment of PV panels (thus without PV powers) as
reference to calibrate the OU parameters of the masked load: 
\begin{align}
\min_{\Theta} || S(P^{\rm obs}_{\rm NET}) - S(P(\Theta)) ||_2 + || \Theta_{\rm
prior} - \Theta ||_2 ,
\label{eq:ML_OU}
\end{align}
where $S(\cdot)$ is a statistic of the net power time series of observed data
and of data generated through simulations by using parameters
$\Theta$. The first term estimates the discrepancy between the
statistics observed and the one generated by the simulated
process. The second term represents a regularization, where $\Theta_{\rm
prior}$ can be either nominal values or zero. The statistic $S$ is
defined by considering the time series that generates a stochastic process
$\hat{X}(\Theta) = \{\hat{x}_t(\Theta)\}$ by $P_{NET}$ via \eqref{eq:ou_em} and
$P_{\rm PV}$ in \eqref{eq:power_balance} and computes a series of statistics such as mean, standard deviation, and weighted autocorrelations' norm:
\begin{align}
S(\hat{X}) := \left[ \mu_{\hat{X}}, \sigma_{\hat{X}}, \frac{1}{t_1}||R_{\hat{X}\hat{X}}(\tau)||_2 \right], \;\; \tau=1,\cdots,t_1 .
\label{eq:ML_ou_d}
\end{align}

\section{Numerical Results\label{sec:numerics}}

We present two examples: a synthetic example
(\S\ref{sec:numerics:synthetic}) and a realistic one (\S
\ref{sec:numerics:realistic}). In the synthetic example we generate the
solar irradiance with known spatial distribution and consider one of
the Oahu island measurements for the temporal correlation. We also
generate a simplified OU masked load power. In the realistic case we
use real $\mu$PMU measurements and irradiance to generate the net
power. The total compute time in all our examples takes a few minutes
on a regular laptop.

\subsection{Synthetic Example\label{sec:numerics:synthetic}}

We start the numerical illustrations of the proposed framework by
using a synthetic example. The point of this example is to test the
framework in ideal situations that correspond to good parametric
modeling of the irradiance, PV, and masked load. To this end, we generate the true masked
load by using an OU process and the true PV power generation by 
using a Gaussian process, both with known parameters. This data set is
used to generate the net load data. The observables in this system are
the net load data and the irradiance at the two locations indicated in
Fig.~\ref{fig:solar:sites}. In this setup we assume that  the Gaussian
process has an exact spatial structure and that the masked load is described
by the correct OU process, but with unknown parameters. We aim to (i)
recover the GP parameters from data and (ii) recover the OU process
parameters of the masked load that together with the PV power best
explain the observed net load.

\subsubsection{Calibrating the GP} We assume that we have 17 PV panels in
a limited area that corresponds to the Oahu irradiance sensor network
(Fig.~\ref{fig:solar:sites}), out from we pick two  as
observations. The exact GP model has the following parameters:
$\alpha=0.0108$, $\beta=0.0001$, $\theta_x=61.6522$,
$\theta_y=74.081$. We first calibrate the GP model by using a least
squares fit and one hour's worth of data (3600 seconds). The resulting
GP parameters are $\alpha=0.01085$, $\beta=1.01e-05$,
$\theta_x=64.44631$, and $\theta_y=70.899$, which is an excellent fit, as
expected. Then we use the data from the two observation sites, the
calibrated GP model and the conditional distribution
\eqref{eq:conditional:GP}, to predict the clear sky 
index at the remaining 15 sites.

\subsubsection{Masked load and PV disaggregation} The disaggregation
problem follows the steps described in \S
\eqref{sec:Disaggregation:Algrithm}. In particular, we solve an
optimization problem that yields the maximum likelihood of the
OU parameters (that define the masked load) that best explain the data
(the net load). The likelihood is expressed in terms of the statistics of the observed and simulated data. 

\begin{figure}[!h]
\centering
\includegraphics[width=1\linewidth]{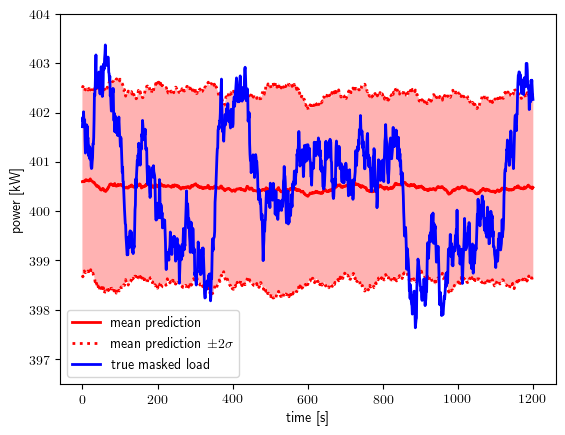}
\caption{Masked load (truth) and its reconstruction (prediction) through our
  disaggregation strategy. The predicted masked load is represented by
  the mean value process and $\pm 2 \sigma$ deviation. The true value
  is covered as close to 90\% by the prediction envelope, as expected.}
\label{fig:net:load:forecast}
\end{figure}

We set the true OU parameters  $\mu_{\rm OU}$, $\gamma_{\rm OU}$, and
$\sigma_{\rm OU}$ to be $[400000, 0.01, 200]$.  Note that in this case
we do not use the jump process. We performed 50 solves 
with initial guesses initialized at $\pm60\%$ around the true
values. The results from solving these problems came to $\mu_{\rm OU}^*
\in (400463.6,400463.8)$,  $\sigma_{\rm OU}^* \in (156.1, 156.8)$, and
$\gamma_{\rm OU}^* \in (0.0128, 0.0131)$. These results indicate that
the minimizer is found closer to the true solution, and thus the
estimator approximates the true values. This results in a 
good representation of the masked load process. In
Fig.~\ref{fig:net:load:forecast} we illustrate the masked load
reconstruction from one such optimization (results look  similar
for the other ones). 

\subsection{Realistic Example\label{sec:numerics:realistic}}

We also tested our framework on real data sets. We used the
same solar PV sites as in the synthetic example, only now the solar
data set is the one actually collected from the sensor network
consisting of the 17 pyranometers, measuring GHI with a $1$ Hz
resolution \cite{Sengupta_2010}. The masked load power measurements are collected
by $\mu$PMUs and PQube3 power quality meters manufactured by Power Standards Laboratory in Alameda,
CA, at $120$ Hz \cite{peisert2017lbnl}. These measurements are downsampled
to $1$ Hz to match the solar sampling rate. The net load power data is
obtained by \eqref{eq:power_balance}.

\subsubsection{Calibrating the GP} 
\begin{figure}[!h]
\centering
\includegraphics[width=1\linewidth]{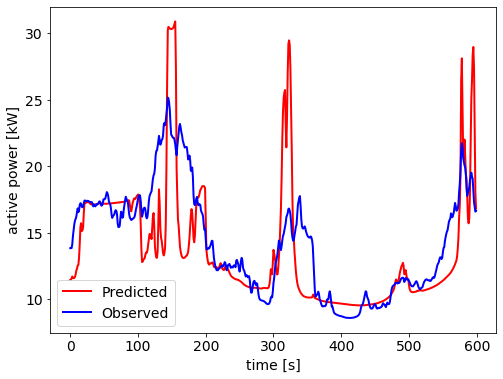}
\caption{Active PV power generation and its prediction through our
  GP model using eq. (\ref{eq:ac_power}). The predicted PV power is inferred by GHI observations on site DH4 and AP5. The true value
  is calculated by full observation of 17 sites.}
\label{fig:pv_power:forecast}
\end{figure}

We assume that among the 17 PV panels we have the observable set $\{DH4, AP5\}$
(Fig. \ref{fig:solar:sites}). We first calibrate the GP model by using a least
squares fit and ten minutes  of data (600 seconds). The resulting
GP parameters are $\alpha=0.09243$, $\beta=1.00e-03$,
$\theta_x=20.14$, and $\theta_y=17.63$, and the estimation error is $|| \Sigma_{model\_opt} - \Sigma_{obs} ||_2=0.223$, which is an excellent fit. Then we use the data from the two observation sites, the
calibrated GP model and the conditional distribution
\eqref{eq:conditional:GP}, to predict the clear sky 
index at the remaining 15 sites and compute the AC PV power prediction. Comparing with the PV power computed by full observation data (all 17 locations) in Fig.~\ref{fig:pv_power:forecast}, our prediction is  close to the true value and successfully predicts the sudden injection jumps.

\subsubsection{Masked load and PV disaggregation} The disaggregation
problem follows the steps described in \S
\eqref{sec:Disaggregation:Algrithm}. We get a rough estimation of OU process parameter set $\Theta_1$. Then we solve an
optimization problem that yields the maximum likelihood of the
OU parameters for $\Theta_{opt}$, taking $\Theta_1$ as the initial.

\begin{figure*}[!ht]
\centering
\includegraphics[width=0.8\linewidth]{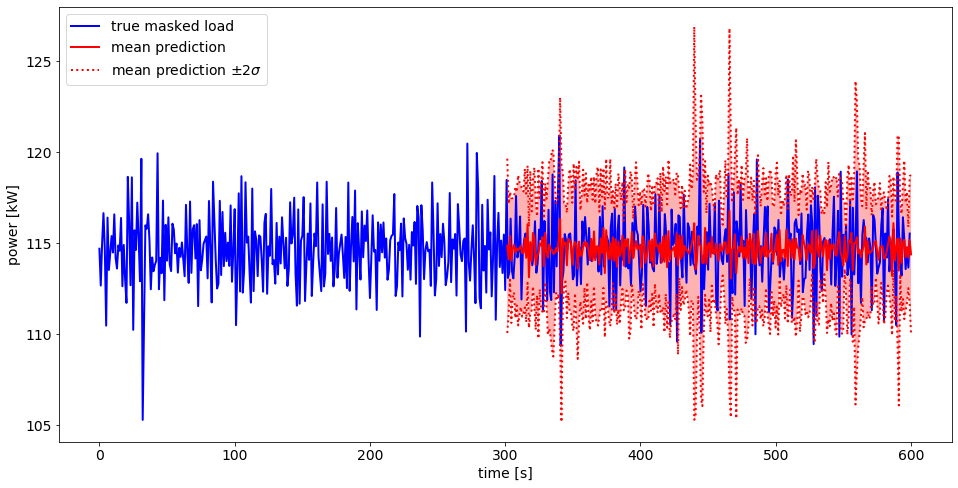}
\caption{Real load power data vs. BTM real power prediction by the Euler-Maruyama scheme for the OU process using parameters calibrated by maximum likelihood. We used 5 minutes of net load and irradiance data for the estimation and make predictions of the masked load for the next 5 minutes. The predicted masked load is represented by the mean value process and $\pm 2 \sigma$ deviation calculated by 10 realizations. The true value is covered as close to 96.67\% by the prediction envelope, as expected.}
\label{fig:load:ou_ml_pred}
\end{figure*}

In the numerical experiment we use 5 minutes of irradiance data and net load data generated by recorded PMUs for disaggregation and parameter estimation. Then we make predictions of the masked load for the next 5 minutes using the estimated parameters. The absolute error of the parameter set suggests the parameter estimation of the OU process is significantly improved by introducing maximum likelihood, compared with the rough estimation in Table \ref{table:2}.

 \begin{table}[h!]
\centering
\begin{tabular}{l|l l l l} 
 \hline
  & $\mu$ & $\gamma$ & $\mu_1$ & $\sigma_1$ \\ [0.5ex] 
 \hline\hline
  $|\Theta_{ref} - \Theta_1|$ & 5.30e+02& 1.97e-01 & 5.74e+01 & 1.53e+03\\
  $|\Theta_{ref} - \Theta_{opt}|$ & 1.58e+01 & 8.39e-02 & 2.46e-02 & 2.11e-02 \\
  \hline
  & $k$ & $\theta$ & $\lambda$ &  \\ [0.5ex] 
 \hline\hline
  $|\Theta_{ref} - \Theta_1|$ & 1.45e-02 & 1.13e+03 & 0.00e+00 &  \\
  $|\Theta_{ref} - \Theta_{opt}|$ & 5.64e-01 & 3.51e-02 & 0.00e+00 & 
 \\ [1ex] 
 \hline 
\end{tabular}
\caption{OU Parameter Estimation by Maximum Likelihood}
\label{table:2}
\end{table}
The predicted masked load is represented by the mean value process and $\pm 2 \sigma$ deviation calculated by 10 random realizations. The true value is covered as close to 96.67\% by the prediction envelope. These intuitive results indicate that
the minimizer is found close to the true solution, which results in a
correct representation of the masked load process. In
Fig.~\ref{fig:load:ou_ml_pred} we illustrate the masked load
reconstruction from one such optimization (results look  similar
for the other ones). Numerical results indicate that our approach can disaggregate the masked load well from the net load using limited irradiance observation. It also capture its trend as well as the variability thus provide a more accurate prediction of the masked load than naive predictions by just assuming the mean and standard deviation from the historical data.

\section{Conclusion and Future Work}
\label{sec:conclusion_and_future_work}

In this paper we present a probabilistic analysis framework to
estimate behind-the-meter photovoltaic generation in a single feeder
network in real time. Within this framework we develop a forward model
consisting of a spatial stochastic process that estimates the
photovoltaic generation based on a couple of sensors and a temporal stochastic differential equation with jumps that estimates the masked user load demand. These
models are used to disaggregate the behind-the-meter photovoltaic
generation by using net load and partial irradiance measurements. Simulation
studies with both synthetic and real recorded solar irradiance data
and $\mu$PMU data indicate that the proposed framework is a tenable method
to provide a reliable disaggregation procedure. Moreover, simulations indicate that we can accurately estimate the aggregated PV active power generation at a distribution feeder with limited sensor deployment. This model takes full
consideration of major characteristics of masked load and PV
production and thus leads naturally to predictive capability in
real time. For larger areas and same density of observations, we
expect this strategy to perform similarly and arguably better with
more sophisticated models that can take advantage of more information. Nevertheless, the results presented in this study are
limited by the availability of measurements and future studies should address larger
areas if data becomes available. 

Our novel framework can be naturally extended to several other
directions, which we plan to investigate. For the solar
generation predicted by partial irradiance measurements, improvements
could be made such that we can accurately predict the irradiance jumps. On the
masked load model and disaggregation side, further improvements could
be made to the SDE model and computational framework. 
Variability on different time horizons could also be considered in the
future for real applications.

\bibliographystyle{IEEEtran}

\end{document}